\documentclass[useAMS,usenatbib,usegraphicx]{mn2e}
\usepackage{epsf}
\title[Local Group Potential Shape]
{The Three Dimensional Structural Shape of the Gravitational Potential 
in the Local Group} 
\author[Bomee Lee and Jounghun Lee]
{Bomee Lee\thanks{E-mail: bmlee@astro.snu.ac.kr} and Jounghun Lee\thanks
{E-mail: jounghun@astro.snu.ac.kr}\\
Department of Physics and Astronomy, FPRD, Seoul National University, 
Seoul 151-747, Korea \\}

\begin{document}

\date{Accepted 2008 ???. Received 2008 ???; in original form 2007 October 10}

\pagerange{\pageref{firstpage}--\pageref{lastpage}} \pubyear{2008}

 \maketitle

\label{firstpage}

\begin{abstract}
The Local Group is a small galaxy cluster with the membership of $62$ nearby 
galaxies including the Milky Way and M31. Although the Local Group has yet to 
be virialized, it interacts with the surrounding matter as one gravitationally 
bound system.  To understand the formation and evolution of the Local Group as 
well as its member galaxies,  it is important to reconstruct the gravitational 
potential field from the surrounding matter distribution in the local cosmic 
web. By measuring the anisotropy in the spatial distribution of the Local 
Group galaxies, which is assumed to be induced by the local gravitational 
tidal field , we resolve the three dimensional structure of the gravitational 
potential in the vicinity of the Milky Way smoothed on the Local Group mass 
scale.  Our results show that (i) the minor principal axis of the Local Group 
tidal field is in the equatorial direction of $\alpha_{p}=15^{h}00^{m}$ and 
$\delta_{p}=20^{d}$; (ii) it has a prolate shape with axial ratio of 
$0.5\pm 0.13$; (iii) the global tides in the Local Group is quite strong, 
which may provide a partial explanation for the low abundance of dwarf 
galaxies in the Local Group.
\end{abstract}


\section{Introduction}

Since \citet{hub36} first recognized its existence, the Local Group has 
been widely investigated by various observations 
\citep[for a review, see][and references therein]{mateo98,van00}. 
It is now well known that the Local Group consists of two massive halos 
(the Milky Way and M31), their satellites, and the neighboring galaxies.  
It has been estimated by numerical simulations based on the disk 
formation model that the dark matter halos of the Milky Way and M31 
have virial radius of approximately $200$-$300$ kpc separated by about 
$700$ kpc \citep{kly-etal02,wid-dub05}.  The comparison of the numerical 
results with observational constraints suggested that the Local Group is 
not yet relaxed with two main halos in the merging process. 

Despite that the Local Group has yet to be virialized, it is also true that 
the Local Group behaves as one gravitationally bound system when interacting 
with the surrounding large-scale structure \citep[e.g.,][]{van00}. 
Therefore, to understand the formation and evolution of the Local Group 
under the gravitational influence from the surrounding matter distribution, 
it should be desirable to resolve the gravitational potential field around 
the Local Group.

As pointed out by \citet{spr-etal04}, it is the structure of the 
gravitational potential that reflects more directly the three dimensional 
distribution of dark matter in a given region. Recently, \citet{hay-etal07} 
showed by numerical analysis of the N-body simulation results that the 
potential field is much smoother than the density field and the iso-potential 
surfaces are well approximated by concentric triaxial ellipsoids. 

We attempt here to resolve the three dimensional structural shape of the 
gravitational potential in the regions around the Local Group, assuming that 
its iso-potential surfaces can be well approximated as a triaxial ellipsoid. 
To achieve this goal, we employ the analytic algorithm developed by 
\citet[][hereafter, LK06]{lee-kan06} which allows us to determine the 
eigenvalues of the tidal tensor (defined as the second derivative of the 
gravitational potential) from the anisotropic spatial distribution of the 
galaxies in a given cluster. 

The underlying logic of the LK06 algorithm is as follows: 
The cluster galaxies are observed to be preferentially located near the major 
axes of the parent clusters \citep{kne-etal04,zen-etal05}.  Recent numerical 
analyses based on N-body simulations have shown that this anisotropic 
distribution of the cluster galaxies originate from the local tidal field in 
the parent clusters \citep{atl-etal06}. What \citet{lee-kan06} found is that 
it is possible to determine the eigenvalues of the local tidal field by 
measuring the anisotropy in the spatial distribution of the cluster galaxies. 
Once the eigenvalues of the tidal field are determined, the axial ratios 
of the iso-potential surfaces can be readily calculated 
\citep{bar-etal86,bon-mye96}.

The plan of this paper is as follows. In $\S 2$, the LK06 model is briefly 
overviewed and our theoretical formula based on the LK06 model is presented. 
In $\S 3$ the observational data are described and the anisotropy in the 
spatial distribution of the Local Group member galaxies is measured. Also 
the structural shape of the gravitational potential in the region 
around the Local Group is resolved. In $\S 4$ the results are discussed and 
a final conclusion is drawn. 

\section{Theoretical Background}

\subsection{The gravitational potential profile}

Let $\Phi({\bf r})$ represent the gravitational potential field smoothed 
on the Local Group mass scale. In the neighborhood of the Local Group, 
the Taylor expansion gives 
\begin{equation}
\label{eqn:tay}
\Phi({\bf r}) \approx \Phi(0) + r_{i}\partial_{i}\Phi|_{0} + 
\frac{1}{2}r_{i}r_{j}\partial_{i}\partial_{j}\Phi|_{0},
\end{equation}
where it is assumed that the center of the mass of the Local Group is 
positioned at the origin of ${\bf r}=0$. Supposing that the Local Group 
represents a local peak of the smoothed potential field, the second 
term proportional to $\partial_{i}\Phi|_{0}$ in equation (\ref{eqn:tay})
becomes zero. The tidal shear tensor ${\bf T}$ in the Local Group is defined 
as $T_{ij} \equiv \partial_{i}\partial_{j}\Phi|_{0}$ at the local peak to 
which the third term in equation (\ref{eqn:tay}) is proportional.

In the frame of the principal axes of the tidal shear tensor, 
equation (\ref{eqn:tay}) can be expressed as
\begin{equation}
\label{eqn:pot}
\Phi({\bf r})-\Phi(0)\approx\frac{1}{2} r^{2}_{i}\lambda_{i},
\end{equation}
where, $\lambda_{1},\ \lambda_{2},\ \lambda_{3}$ are the eigenvalues of 
${\bf T}$ with $\lambda_{1}\ge\lambda_{2}\ge\lambda_{3}$. 
Let $a_{1},\ a_{2},\ a_{3}$ be the three semi-axes of an ellipsoidal 
iso-potential surface of $\Phi({\bf r})=const.\equiv \phi$. 
By equation (\ref{eqn:pot}), one can relate $\{ a_{i}\}_{i}^{3}$ to 
$\{ \lambda_{i}\}_{i}^{3}$ as
\begin{equation}
\label{eqn:axes}
a_{i}=\left [ \frac{2(\Phi(0)-\phi)}{\lambda_{i}} \right ] ^{1/2}.
\end{equation}
Therefore, the two axial ratios of the ellipsoidal iso-potential surfaces can 
be written only in terms of $\{ \lambda_{i} \}^{3}_{i=1}$ as 
\begin{equation}
\label{eqn:ratio}
\frac{a_1}{a_3} = \sqrt \frac{\lambda_{3}}{\lambda_{1}}, \qquad
\frac{a_2}{a_3} = \sqrt \frac{\lambda_{3}}{\lambda_{2}}.
\end{equation}
Equation (\ref{eqn:ratio}) implies that once the eigenvalues of the tidal 
shear tensor at the Local Group region are determined, the three dimensional 
shape of the Local Group potential can be reconstructed.      

\subsection{Overview of the LK06 algorithm}

Assuming that the tidal field in the host halo is responsible for the 
anisotropic spatial distribution of the substructures, \citet{lee-kan06} 
have derived an analytic expression for the probability distribution of the 
cosines of the polar angles of the substructure position vectors in the 
tidal shear principal axis frame as 
 \begin{eqnarray}
&p(\cos\theta)=\frac{1}{2\pi} \prod_{i=1}^{3}
(1-s+3s\hat{\lambda}^{2}_{i})^{-1/2} \nonumber \\
&\times  \int_{0}^{2\pi} \left(\frac{\sin^2\theta \cos^2\phi}{1-s+3s
 \hat{\lambda}^{2}_{1}} 
 + \frac{\sin^2\theta \sin^2\phi}
{1-s+3s \hat{\lambda}^{2}_{2}}+\frac{\cos^2\theta}{1-s+3s
\hat{\lambda}^{2}_{3}} \right)^{-3/2} d\phi,
 \label{eqn:pri}
\end{eqnarray}
where $s$ is a correlation parameter in the range of $[-1,1]$ and 
$\{\hat{\lambda}_{i}\}^{3}_{i=1}$ are the unit eigenvalues of 
${\bf T}$ satisfying the constraint of $\hat{\lambda}^{2}_{1}+
\hat{\lambda}^{2}_{2}+\hat{\lambda}^{2}_{3}=1$. Here, the free parameter $s$ 
quantifies the degree of the alignment between the minor principal axis of the 
tidal tensor and the position vectors of the substructures:  If $s=-1$, then 
the substructures are most strongly aligned with the minor principal axis of 
the tidal tensor; If $s=1$, they are most strongly anti-aligned while the case 
of $s=0$ corresponds to no alignment.

Since the minor principal axis of the tidal tensor at the region of a given 
host halo corresponds to the direction in which the tidal force is least 
strong and thus the host halo is most elongated, \citet{lee-kan06} assumed 
that the minor principal axis of the tidal tensor is parallel to the major 
principal axis of the inertia tensor of the host halo. Under this assumption, 
they suggested that by measuring $p(\cos\theta)$ from the spatial distribution 
of the substructure positions and fitting it to equation (\ref{eqn:pri}), 
one can determine the unit eigenvalues of the tidal tensor and the value of 
the correlation parameter, which will in turn yield the values of the axial 
ratios of the host halo. To demonstrate the validity of their algorithm, 
\citet{lee-kan06} tested it against N-body results, and showed that their 
algorithm indeed reconstruct the actual shapes of cluster halos within 
$20\%$ error. 

In the original approach, \citet{lee-kan06} used the virialization 
constraint of $\sum_{i}^{3}\lambda_{i}=\delta_{c}$ where 
$\delta_{c}\approx1.68$ is the density threshold for a virialized structure 
\citep{eke-etal96}. However, note that the LK06 algorithm can be applied 
even to a region that is not completely virialized, since the 
virialization condition was not used to derived equation (\ref{eqn:pri}) 
itself. In fact, it can be applied to any region as long as the member 
objects in the region show tidally induced anisotropy. 
For instance, it can be used to describe the anisotropic spatial distribution 
of galaxy clusters in the superclusters due to the supercluster tidal field 
(i.e., the linear tidal field smoothed on the supercluster mass scale), 
even though the superclusters are not virialized objects \citep{lee-evrard07}. 
It has been also used to describe the anisotropic spatial distribution 
of void galaxies due to the tidal effect from the surrounding filaments 
even though the voids are obviously not virialized \citep{lee-par06}. 

When the LK06 algorithm is applied to an unviriazlied structure, 
the additional constraint of $\sum_{i}^{3}\lambda_{i}=\delta_{c}$ should 
not be used. Therefore, what one can reconstruct by the LK06 algorithm for 
this case is not the shape defined by the inertia momentum tensor of the 
structure but the shape of the potential around the structure.
In the following section, we reconstruct the three dimensional structural 
shape of the gravitational potential in the region around the Local Group 
by measuring the values of $s$ and $\{\hat{\lambda}_{i}\}^{3}_{i=1}$ 
from recent observational data.

\section{Application to the Local Group Data}

\subsection{Tidal Alignment of the Local Group}

We use the observational data of 62 member galaxies belonged to the Local Group.
Table \ref{tab_lg1} lists the latest information on the equatorial coordinate
($\alpha$, $\delta$), distance ($r$) in unit of Mpc and absolute magnitude 
($M_v$) of each member galaxy.  The information on $\alpha$, $\delta$, $r$ 
and $M_{v}$ for the $\# 1-40$ and $\# 48$ galaxies are obtained from 
\citet{kar-etal04}, and for the other galaxies, we show in the seventh columns 
the references from which we obtained the information. For those galaxies 
marked by $\star$ in Table \ref{tab_lg1} we used the updated information 
given in \citet{mcc-etal05}.  Since the mass information is not available, 
we calculate the center-of-luminosity (CL) instead of the center-of-mass (CM) 
for the Local Group by using the information on the absolute magnitude of each 
member galaxies. The Cartesian position of the Local Group CL in the 
equatorial coordinate system is found to be ${\rm CL}=[0.351, 0.069, 0.259]$ 
in unit of Mpc.

\begin{figure}
\begin{center}
\includegraphics[width=1.0\hsize]{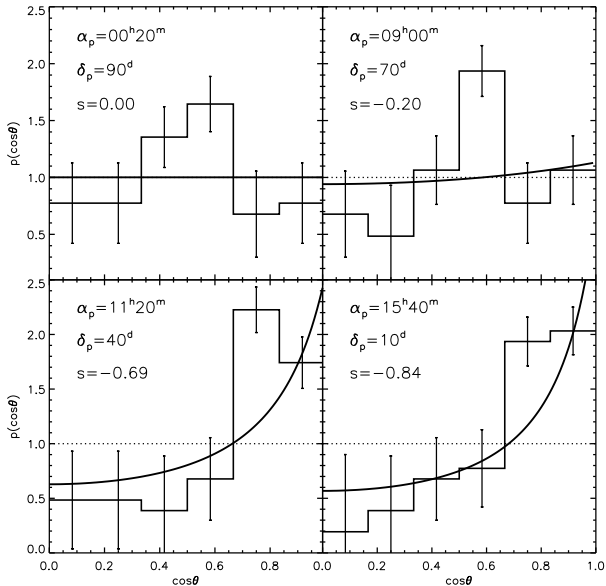}
\end{center}
\caption{Probability density distribution of the cosines of the angles 
between the position vectors of the Local Group member galaxies in the CL 
frame and four different choices of major principal axis of the Local Group. 
In each panel, the histogram with Poisson errors represents the observational 
data points of Local Group, the solid line is the analytic fitting from LK06 
reconstructing algorithm, and the dotted line corresponds to the case of 
no correlation. Four plots in each panel show how the agreements between the 
observational and the analytic results change if different directions are used 
rather than the major principal axis.}
\label{fig:pa}
\end{figure}

We measure the three dimensional positions of the Local Group member galaxies 
in the CL frame in the equatorial coordinates. Then, we attempt to find a 
direction in the equatorial coordinates, say, $\alpha_{p}$ and $\delta_{p}$, 
with which the position vectors of the Local Group member galaxies in the 
CL frame show the highest degree of anisotropy. We will regard this direction, 
if found, as the direction of the minor principal axis of the local tidal field.
\begin{table*}
\begin{minipage}{130mm}
\caption{Observational Data on the Local Group}\label{tab_lg1}
\begin{tabular}{ccccccc}
\hline
No & Name &
\multicolumn{1}{c}{$\alpha(2000)$} &
\multicolumn{1}{c}{$\delta(2000)$} &
\multicolumn{1}{c}{$r$} & 
\multicolumn{1}{c}{$M_v$} &
Reference \\
&&
\multicolumn{1}{c}{$[h:m:s]$} & 
\multicolumn{1}{c}{$[d:m:s]$} & 
\multicolumn{1}{c}{$[Mpc]$} \\
\hline
1  &  WLM               & 00 01 58.1 & -15 27 40 & 0.932 $^{\star}$  & -14.4 & \\
2  &  NGC55             & 00 08 13.3 & -34 13 13 & 1.80  & -17.5 & \\
3  &  IC 10             & 00 20 24.5 &  59 17 30 & 0.66  & -10.1 & \\
4  &  NGC147            & 00 33 11.6 &  48 30 28 & 0.675 $^{\star}$ & -15.1 &\\
5  &  And III           & 00 35 33.8 &  36 29 52 & 0.749 $^{\star}$ & -10.2 &\\
6  &  NGC 185           & 00 38 58   &  48 20 10 & 0.616 $^{\star}$ & -15.6 &\\
7  &  NGC 205           & 00 40 22.5 &  41 41 11 & 0.824 $^{\star}$ & -16.4 &\\
8  &  M32               & 00 42 42.1 &  40 51 59 & 0.77  & -16.5 &\\
9  &  M31               & 00 42 44.5 &  41 16 09 & 0.785 $^{\star}$ & -21.2 &\\
10 &  And I             & 00 45 40   &  38 02 14 & 0.745 $^{\star}$ & -11.8 &\\
11 &  SMC               & 00 52 38   & -72 48 01 & 0.06  & -17.1 &\\
12 &  Sculptor          & 01 00 09.4 & -33 42 33 & 0.09  & -9.8  &\\
13 &  LGS 3             & 01 03 56.6 &  21 53 41 & 0.769 $^{\star}$ & -10.4 &\\
14 &  IC 1613           & 01 04 54.10&  02 07 60 & 0.73  & -15.3 &\\
15 &  And II            & 01 16 29.8 &  33 25 09 & 0.652 $^{\star}$ & -9.1  &\\
16 &  M33               & 01 33 50.8 &  30 39 37 & 0.809 $^{\star}$ & -18.9 &\\
17 &  Phoenix           & 01 51 06.3 & -44 26 41 & 0.44  & -9.8  &\\
18 &  Fornax            & 02 39 54.7 & -34 31 33 & 0.14  & -13.1 &\\
19 &  EGB 0427+63       & 04 32 00.3 &  63 36 50 & 1.80  & -15.9 &\\
20 &  LMC               & 05 23 34.6 & -69 45 22 & 0.05  & -18.5 &\\
21 &  Carina            & 06 41 36.7 & -50 57 58 & 0.10  & -9.4  &\\
22 &  Leo A             & 09 59 26.4 &  30 44 47 & 0.69  & -11.5 &\\
23 &  Sextans B         & 10 00 00.1 &  05 19 56 & 1.36  & -14.0 &\\
24 &  NGC 3109          & 10 03 07.2 & -26 09 36 & 1.33  & -15.5 &\\
25 &  Antlia            & 10 04 04   & -27 19 55 & 1.32  & -9.8  &\\
26 &  Leo I             & 10 08 26.9 &  12 18 29 & 0.25  & -11.9 &\\
27 &  Sextans A         & 10 11 00.8 & -04 41 34 & 1.32  & -13.9 &\\
28 &  Sextans           & 10 13 03   & -01 36 52 & 0.09  & -9.5  &\\
29 &  Leo II            & 11 13 29.2 &  22 09 17 & 0.21  & -10.1 &\\
30 &  GR 8              & 12 58 40.4 &  14 13 03 & 2.10  & -12.0 &\\
31 &  Ursa Minor        & 15 09 11.3 &  67 12 52 & 0.06  & -8.9  &\\
32 &  Draco             & 17 20 01.4 &  57 54 34 & 0.08  & -8.6  &\\
33 &  Milky Way         & 17 45 09.2 & -28 01 12 & 0.01  & -20.9 &\\
34 &  Sagittarius       & 18 55 03.1 & -30 28 42 & 0.02  & -13.8 &\\
35 &  SagDIG            & 19 29 59   & -17 40 41 & 1.04  & -12.0 &\\
36 &  NGC 6822          & 19 44 57.7 & -14 48 11 & 0.50  & -16.0 &\\
37 &  Aquarius          & 20 46 51.8 & -12 50 53 & 1.071$^{\star}$ & -10.9 &\\
38 &  IC 5152           & 22 02 41.9 & -51 17 43 & 2.07  & -15.6 &\\
39 &  Tucana            & 22 41 49   & -64 25 12 & 0.88  & -9.6  &\\
40 &  UKS 2323-326      & 23 26 27.5 & -32 23 26 & 2.23  & -12.9 &\\
41 &  Pegasus           & 23 28 34.1 &  14 44 48 & 0.919$^{\star}$ & -12.3 & \citet{mateo98}\\
42 &  And VII           & 23 26 31.8 & +50 40 32 & 0.763$^{\star}$ & -13.3 & \citet{van00}\\
43 &  And VI            & 23 51 46.4 & +24 35 10 & 0.82  & -11.5 & \citet{whi-etal07}\\
44 &  Cetus             & 00 26 11.0 & -11 02 40 & 0.755$^{\star}$ & -11.3 & \citet{whi-etal99}\\
45 &  And V             & 01 10 17.1 & +47 37 41 & 0.774$^{\star}$ & -9.6  & \citet{van00}\\
46 &  And IX            & 00 52 52.8 & +44 12 00 & 0.765$^{\star}$ & -8.3  & \citet{zuc-etal04}\\
47 &  Canis Major dwarf & 07 12 36   & -27 40 00 & 0.008 &   -   & \citet{mar-etal04}\\
48 &  Cas1              & 02 06 07.9 & +69 00 36 & 3.3   & -15.6 & \\
49 &  Bootes dwarf      & 14 00 06   & +14 30 00 & 0.06  & -5.8  & \citet{bel-etal06}\\
50 &  Canes Ven
atici    & 13 28 03.5 & +33 33 21 & 0.22  & -7.9  & \citet{zuc-etal06a}\\
51 &  Willman1          & 10 19 22   & +51 03 03 & 0.045 & -3.0  & \citet{wil-etal05b}\\
52 &  Ursa Major        & 10 34 53   & +51 55 12 & 0.1   & -6.75 & \citet{wil-etal05a}\\
53 &  Ursa Major II     & 08 51 30   & +63 07 48 & 0.032 & -3.8  & \citet{zuc-etal06b}\\
54 &  Heracules         & 11 32 57   & -00 32 00 & 0.16  & -5.1  & \citet{bel-etal07}\\
55 &  Coma Berenices    & 12 26 59   & +23 54 15 & 0.044 & -3.7  & \citet{bel-etal07}\\
56 &  Segue1            & 10 07 04   & +16 04 55 & 0.023 & -3.0  & \citet{bel-etal07}\\
57 &  Leo IV            & 16 31 02   & +12 47 30 & 0.14  & -6.0  & \citet{bel-etal07}\\
58 &  And X             & 01 06 33.7 & +44 48 15.8 & 0.7   & -8.1& \citet{zuc-etal07} \\

\end{tabular} 
\end{minipage}
\end{table*}

\begin{table*}
\begin{minipage}{120mm}
\begin{tabular}{ccccccc}
59 &  Canes Venatici II & 12 57 10   & +34 19 15 & 0.15  & -4.8  & \citet{bel-etal07}\\
60 &  Bootes II         & 13 58 00   & +12 51 00 & 0.06  &   -   & \citet{wal-etal07}\\
61 &  Leo T             & 09 34 53.4 & +17 03 05 & 0.42  & -7.2  & \citet{rya-etal08}\\
62 &  And XVII          & 00 37 07   & +44 19 20 & 0.794 & -8.5  & \citet{irw-etal08}\\
\hline
\end{tabular} 

{\footnotesize} References: No.$1 \sim  40$ and 48 from \citet{kar-etal04}; 
 $^{\star}$~\citet{mcc-etal05}
\end{minipage}
\end{table*}
 
Let $\theta$ represent the angle between the equatorial position vector of 
a given member galaxy and the direction of ($\alpha_{p}$, $\delta_{p}$). 
Changing the values of $\alpha_{p}$ and $\delta_{p}$ by $5^\circ$ 
systematically from $0^\circ$ to $360^\circ$ and $-90^\circ$ to $90^\circ$, 
respectively, we measure repeatedly the probability distribution, 
$p(\cos\theta)$, and search for ($\alpha_{p}$, $\delta_{p}$) 
which leads to the maximum deviation of $p(\cos\theta)$ from uniform 
distribution. The bin size of $5^\circ$ is chosen as the smallest bin size 
below which the numerical noise dominate.
By fitting equation (\ref{eqn:pri}) to the observational results, 
we determine a direction of ($\alpha_{p}$, $\delta_{p}$) which leads 
the maximum deviation of $p(\cos\theta)$ from uniform distribution. 

\begin{figure}
\begin{center}
\includegraphics[width=1.0\hsize]{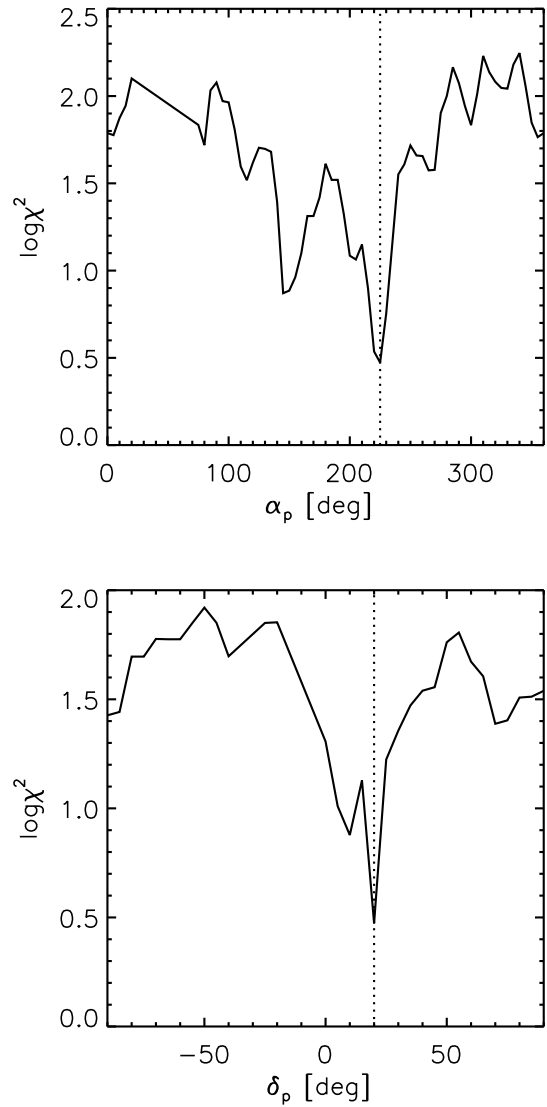}
\end{center}
\caption{The variation of $\chi^2$ as a function of $\alpha_{p}$ (top) and 
$\delta_{p}$ (bottom) when the other parameters are fixed at the best-fit 
values. In each panel the vertical dotted line indicates the position at 
which $\chi^2$ has the minimum value 
($\alpha_{p}=225^\circ$, $\delta_{p}=20^\circ$).}
\label{fig:radec}
\end{figure}
\begin{figure}
\begin{center}
\includegraphics[width=1.0\hsize]{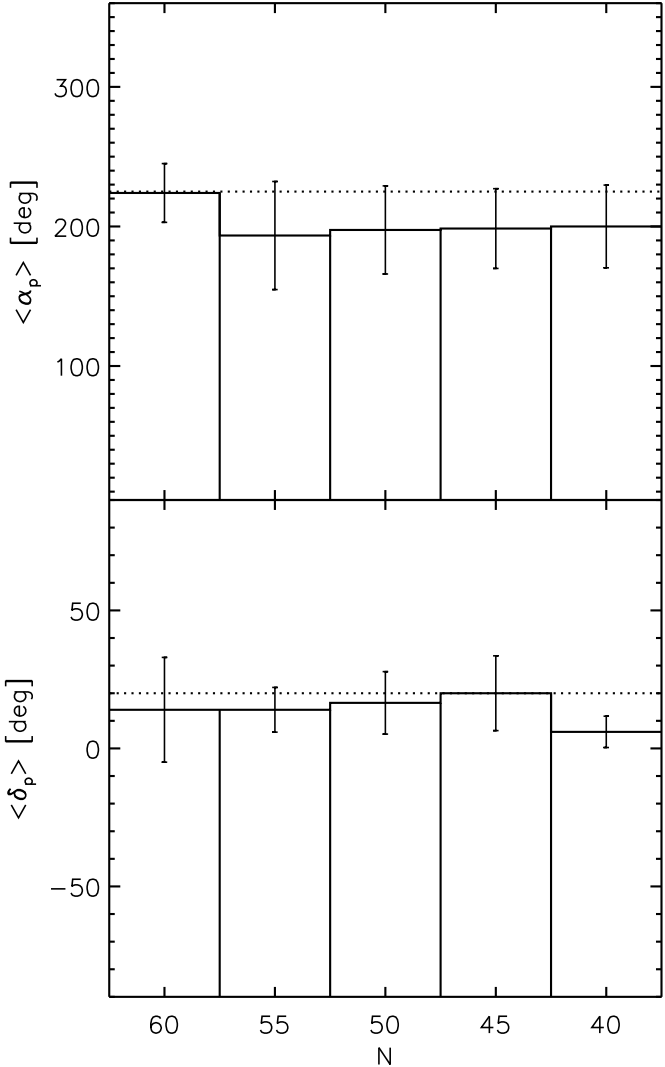}
\end{center}
\caption{The mean values, $\langle\alpha_p\rangle$ (top) and 
$\langle\delta_p\rangle$ (bottom), averaged over $10$ samples of $N$ randomly 
selected galaxies out of the $62$ Local Group members. In each panel the 
errors represent one standard deviation between samples and the horizontal 
dotted line correspond to our result of 
($\alpha_{p}=225^\circ$, $\delta_{p}=20^\circ$).}
\label{fig:ran}
\end{figure}

Fig. \ref{fig:pa} plots the cases of four different directions of the minor 
principal axis of the Local Group tidal field as examples. 
In each panel, the histogram with Poisson errors corresponds to the 
observational results while the solid curve is the fitting model 
(eq.[\ref{eqn:pri}]) with the best-fit values of $\hat{\lambda}_{1}$, 
$\hat{\lambda}_{2}$ and $s$: For a given direction of 
($\alpha_{p}$, $\delta_{p}$), we fit equation (\ref{eqn:pri}) to the 
observational result of $p(\cos\theta)$ by adjusting the values of 
$\hat{\lambda}_{1}$, $\hat{\lambda}_{2}$, and $s$ and determine for the 
best-fit  values of the three parameters with the help of the $\chi^{2}$ 
statistics. As can be seen in Fig. \ref{fig:pa}, the observed distribution 
of $p(\cos\theta)$ behaves quite differently as the direction of 
($\alpha_{p}$, $\delta_{p}$),changes. For a certain direction of 
$(\alpha_{p},\delta_{p})$, the distribution becomes quite uniform (top-left). 
While for another direction $p(\cos\theta)$ increases sharply with 
$(\cos\theta)$ indicating preferential alignments of the galaxy's position 
vectors with the given direction and the analytic fits work well 
(bottom-right).
 
It is found that the direction of 
($\alpha_{p}=15^{h}00^{m}$, $\delta_{p}=20^{d}$) 
leads to the maximum alignments. In other words, when this direction is 
chosen as the minor principal axis of the Local Group tidal field, 
equation (\ref{eqn:pri}) agrees with the observational results best 
with the value of $\chi^{2}$ minimized.  Fig. \ref{fig:radec} plots 
$\chi^{2}$ as a function of $\alpha_{p}$ (top) and $\delta_{p}$ (bottom) 
when the three other parameters are fixed at their best-fit values. 
In each panel the vertical dotted line indicates the position at 
which $\chi^2$ has the minimum value. It is worth mentioning here that 
the direction we found as the minor principal axis of the Local Group 
tidal field $(\alpha=15^{h}00^{m}$, $\delta=20^{d})$ does not coincide with 
the direction parallel to the connection line between M31 and MW. 
It indicates that the observed anisotropic distribution of the LG 
galaxies is induced by the tidal effect of the surrounding matter distribution 
not by the fact that the majority of them are the satellites of M31 and MW.

To examine how stable our result is, we construct $10$ samples of $N$ 
randomly selected galaxies out of the parent sample of $62$ members. 
For each sample, we determine the best-fit values of $\alpha_p$ and 
$\delta_p$ using the same procedure described in the above, and calculate 
the means of $\alpha_p$ and $\delta_p$ averaged over $10$ samples. 
Fig.~3 plots $\langle\alpha_p\rangle$ and $\langle\delta_p\rangle$ as a 
function of $N$ as histogram in the top and bottom panel, respectively. 
In each panel the errors represent one standard deviation between samples 
and the horizontal dotted line represents our best-fit result
($\alpha_{p}=225^\circ$, $\delta_{p}=20^\circ$), obtained from the $62$ LG 
galaxies. As can be seen,  the variations of $\langle\alpha_p\rangle$ and 
$\langle\delta_p\rangle$ with $N$ stay within one standard deviation, 
until the value of $N$ decreases down to $45$. At the bin of $N=40$, however, 
$\langle\delta_p\rangle$ suddenly drops by more than two standard deviations.
Therefore, one can say that our result is stable as long as more than 
$45$ LG galaxies are used to reconstruct the minor principal axis of 
the tidal field.

Since the effect of the nonlinear process like merging between MW and M31 
should be dominant near the LG center, the spatial distribution of those 
galaxies located near the LG center like the satellites of the MW and M31 
are not good indicators of the tidal effect. Only when a large fraction of 
distant member galaxies whose positions are not severely modified by the 
merging effect are included, the overall spatial distribution of the LG 
member galaxies will reveal the anisotropy induced by the local tidal effect.
\begin{figure*}
\begin{center}
\includegraphics[width=1.0\hsize]{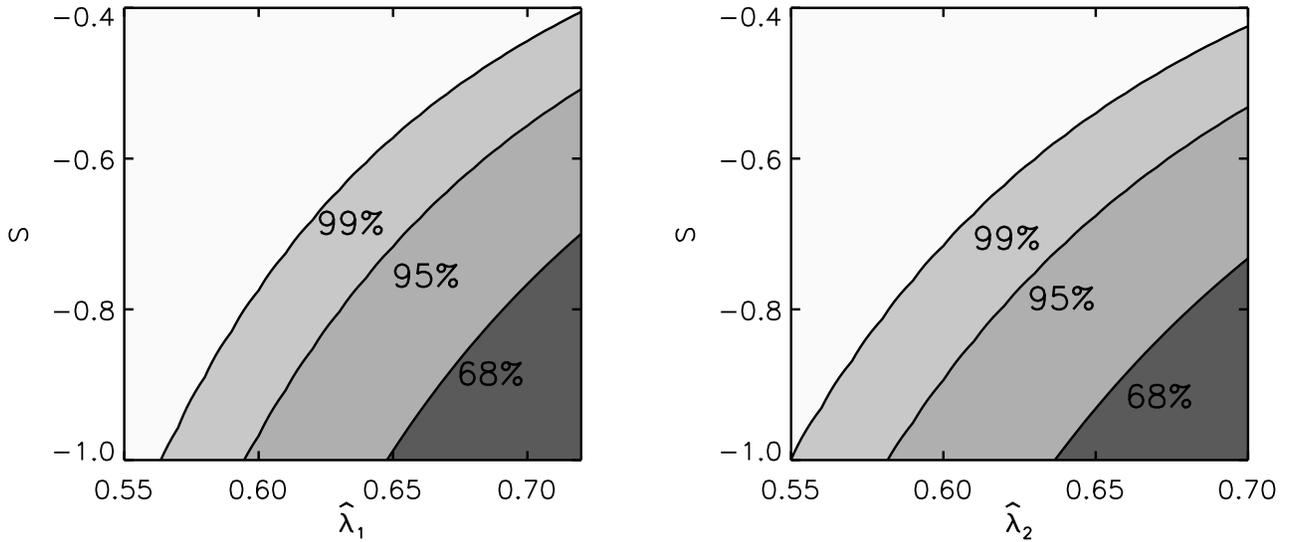}
\end{center}
\caption{$65\%$, $95\%$ and $99\%$ contour plots for $s$ and 
$\hat{\lambda}_{1}$ (left) and for $s$ and $\hat{\lambda}_{2}$ (right).
In each panel the other parameter is fixed at its best-fit value.}
\label{fig:cont}
\end{figure*} 

\begin{figure}
\begin{center}
\includegraphics[width=1.0\hsize]{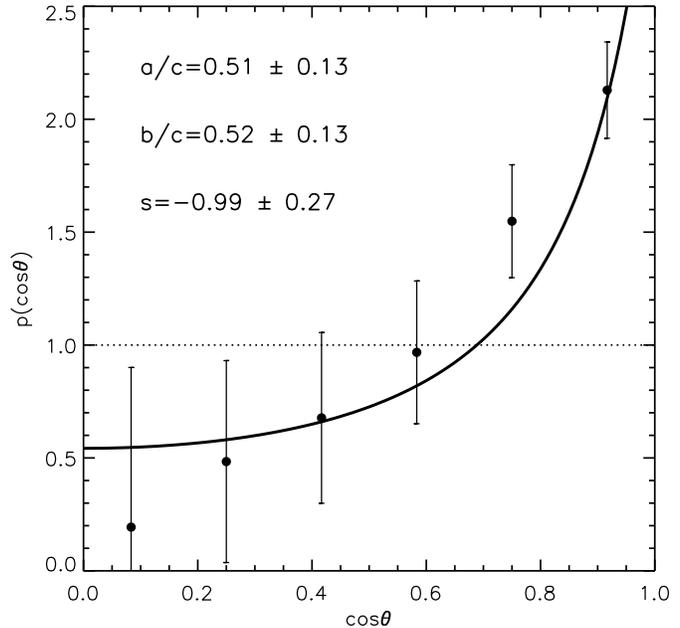}
\end{center}
\caption{Probability distribution of the cosines of the angles between the 
spatial positions of the Local Group member galaxies and the major principal 
axis, $\alpha=15^{h}00^{m} $, $\delta=20^{d}$. The solid line corresponds to 
the fitting model (eq.[\ref{eqn:pri}]) with the best-fit parameters of 
$s=-0.99$ and $a_{1}/a_{3}=0.51$, $a_{2}/a_{3}=0.52 $ while the solid dots 
with Poisson errors represent the observational results.}
\label{fig:best}
\end{figure}

When the minor principal axis of the Local Group tidal field is determined 
as this direction, the values of the fitting parameters are found to be 
$\hat{\lambda}_{1}=0.7\pm 0.04$, $\hat{\lambda}_{2}=0.69 \pm 0.04$ and 
$s=-0.99 \pm 0.27$. Note that this corresponds to the unit eigenvalues of the 
tidal field smoothed on the Local Group mass scale. 
Fig. \ref{fig:best} compares this best-fit model (solid line) with the 
observational result (dots) with Poisson errors for the minor axis direction 
of ($\alpha_{p}=15^{h}00^{m}$, $\delta_{p}=20^{d}$).
As can be seen, the fitting model agrees with the observational result pretty 
well. The probability distribution of $p(\cos\theta)$ with the best-fit 
parameters indeed shows strong alignments between the minor axis of the Local 
Group tidal field and the positions of the Local Group galaxies. 

The unmarginalized error in the measurements of the best-fit values of 
$\hat{\lambda}_{1}$, $\hat{\lambda}_{2}$ and $ s$ are calculated by using 
the formula given in \citep{bev-rob96}. For instance, the error in the 
measurement of $\hat{\lambda}_{1}$ is calculated as 
\begin{equation}
\label{eqn:err}
\sigma_{\hat{\lambda}_{1}}=\Delta \hat{\lambda}_{1}\sqrt{2(\chi_{1}^{2}-
2\chi_{2}^{2}+\chi_{3}^{2})^{-1}}.
\end{equation}
where $\Delta \hat{\lambda}_{1} $ is the bin size of $\hat{\lambda}_{1}$.
Here, $\chi_{i}^{2}=\chi^{2}(\hat{\lambda}_{1i})|_{i=1,2,3}$ and 
$\hat{\lambda}_{12}=\hat{\lambda}_{11}+\Delta \hat{\lambda}_{1}$, 
$\hat{\lambda}_{13}=\hat{\lambda}_{12}+\Delta \hat{\lambda}_{1}$ and 
$\hat{\lambda}_{11}$ is the best-fit value. The error associated with the 
best-fit values of $\hat{\lambda}_{1}$ and $s$ are also calculated in a 
similar manner. To account for the parameter degeneracy and estimate the 
marginalized errors in the measurement of the best-fit values, we also 
compute the $68\%$, $95\%$ and $99\%$ confidence regions for 
$(s,\ \hat{\lambda}_{1})$ and for $(s,\ \hat{\lambda}_{2})$, 
which are shown in  in the left and right panels of Fig. \ref{fig:cont}, 
respectively. 

By expressing equation (\ref{eqn:ratio}) in terms of the unit eigenvalues, 
we can now measures the three dimensional shape of the gravitational 
potential in the region around the Local Group: 
\begin{equation}
\label{mod}
a_{1}/a_{3}=\sqrt\frac {\hat{\lambda}_{3}}{\hat{\lambda}_{1}}, \qquad
a_{2}/a_{3}=\sqrt\frac {\hat{\lambda}_{3}}{\hat{\lambda}_{2}}.
\end{equation}
The axial ratios of the ellipsoidal iso-potential surfaces are calculated as 
$a_{1}/a_{3}=0.51\pm 0.13$ and $a_{2}/a_{3}=0.52\pm 0.13$. The associated 
errors are estimated by means of the error propagation method 
\citep{bev-rob96} as
\begin{equation}
\label{eqn:err_ab}
\sigma_{f}^2=\sigma_{\hat{\lambda}_{1}}^{2}\left( 
             \frac{\partial f}{\partial \hat{\lambda}_{1}} \right)^2+
             \sigma_{\hat{\lambda}_{2}}^{2}\left( 
             \frac{\partial f}{\partial \hat{\lambda}_{2}} \right)^2
\end{equation}
where $f$ represents the two axial ratios. Our results indicate that the 
gravitational potential in the Local Group has a prolate shape. Figure 
\ref{fig:best} plots our final result.

\subsection{Strength of the Tidal Effect in the Local Group}

Now that we have determined the unit eigenvalues of the tidal shear field in 
the region around the Local Group, we can estimate the strength of the 
global tides in the Local Group.  Since the global tides plays a role 
of stripping or destroying the substructures, if it is found to be strong 
in the Local Group, then it might explain at least partially the low 
abundance of member galaxies in the Local Group.

The degree of the tidal effect can be quantified in terms of the differences 
between the three eigenvalues of the tidal tensor as \citep{chi-etal01}
\begin{equation}
\label{eqn:r}
\hat{\Lambda}_{\rm T}= \sqrt{\frac{1}{3}
[(\hat{\lambda}_{1}-\hat{\lambda}_{2})^2 + (\hat{\lambda}
_{2}-\hat{\lambda}_{3})^2 + (\hat{\lambda}_{1}-\hat{\lambda}_{3})^2]}. 
\end{equation}
Here the value of $\hat{\Lambda}_{\rm T}$ is in the range of $[0,1]$. 
The case of $\hat{\Lambda}_{\rm T}=0$ corresponds to no tidal effect. 
The higher the degree of the deviation of $\hat{\Lambda}_{\rm T}$ from zero 
is, the stronger the tidal effect is. 
It is found that $\hat{\Lambda}_{\rm T}=0.42$ for the Local Group 
from the best-fit values of $\hat{\lambda}_{1}$ and $\hat{\lambda}_{2}$ 
obtained in \S 3.1, which suggests that the global tides in the Local Group 
is quite strong. 
 
\section{Discussion and Conclusion}

We have shown here that the anisotropic spatial distribution of the Local 
Group galaxies can be used to reconstruct by means of the LK06 algorithm 
the shape of the gravitational potential in the vicinity of the MW smoothed 
on the Local Group mass scale and measure the strength of the tidal effect, 
which is important to understand the interaction of the Local Group as a 
whole with the surrounding matter distribution and its evolution.

In their original approach, \citet{lee-kan06} reconstructed the ellipsoidal 
shape of the host halos rather than the gravitational potential from the 
anisotropic spatial distribution of the halo substructures by using the 
Zel'dovich approximation \citep{zel70} as well as the virialization condition. 
We did not apply their original approach to the Local Group since the 
Local Group has yet to be completely virialized \citep{kly-etal02,wid-dub05}.

Nevertheless, the LK06 algorithm can in fact apply to any overdense region 
in which the embedded structures show anisotropic spatial distribution since 
it has not been derived under a virialization condition. 
For example, \citet{lee-evrard07} used the LK06 algorithm to reconstruct the 
filamentary shapes of the superclusters by measuring the cluster-supercluster 
alignments even though the superclusters are not completely relaxed systems. 
For the case of unrelaxed systems, however, what can be reconstructed by the 
LK06 algorithm is the anisotropy in the gravitational potential but not the 
shape of a bound halo.

Here, we have reconstructed the anisotropy in the gravitational potential 
around the Local Group from the spatial distribution of the Local Group 
galaxies. It is shown that the Local Group potential is prolate with axial 
ratios of $0.5 \pm 0.13$ with the major principal axis parallel to the 
direction of the equatorial coordinate, $\alpha=15^{h}00^{m}$, $\delta=20^{d}$.
Even though our result still suffers from large errors due to small number 
statistic, it has been obtained without resorting to any additional 
assumption on the dynamical state or the shape of the Local Group. 

It is worth discussing one caveat that our result is subject to. 
As mentioned in Section~3.1, the reconstruction procedure is stable as long 
as more than $40$ galaxies of the Local Group are used. We think that this 
caveat comes from the radial dependence of the final result.
Since the effect of such nonlinear process as gravitational merging and 
interaction between the MW and the M31 should be dominant in the inner 
region of the Local Group, the spatial distribution of those nearby 
galaxies like the satellites of the MW and M31 should not be good indicators 
of the tidal effect. 

Meanwhile the positions of the distant galaxies located in the outer region 
of the Local Group are not severely modified by the nonlinear effect and thus 
their spatial distribution will reflect better the effect of the Local Group 
tidal field. It would be desirable to investigate how the triaxial structure 
of the Local Group potential changes with radial distance. It is, however, 
beyond the scope of this paper since the total number of the Local Group 
member galaxies is too small to obtain reliable statistical result on the 
radial dependence. We plan to apply the LK06 algorithm to the Virgo cluster 
and investigate the radial dependence of the Virgo cluster potential 
shape, given the fact that the Virgo cluster has much more member 
galaxies than the Local Group. We wish to report the result in the 
near future.

From the measured eigenvalues of the tidal tensor, we have also shown 
that the the global tides are quite strong in the current Local Group system, 
which may provide an explanation to the low abundance of dwarf galaxies in 
the Local Group \citep{som02}: In the Local Group system where the 
two main galaxies are in the process of filamentary merging the low-mass 
dwarf galaxies are hard to survive due to the strong global tides. 
An interesting question to ask is what the critical mass of the member galaxy 
is to survive the global tides in the Local Group? We also intend to work 
on this issue in the future. 

\section*{Acknowledgments}

We are very grateful to A. Knebe for his very useful referee report 
which helped us improve the original manuscript significantly. 
We acknowledge the Korea Science and Engineering Foundation (KOSEF) grant 
funded by the Korean Government (MOST,NO. R01-2007-000-10246-0).

\bsp

\label{lastpage}

\end{document}